\documentclass[a4paper]{spie}  
\usepackage[]{graphicx}

\title{MESO-NH SIMULATIONS OF THE ATMOSPHERIC FLOW ABOVE THE INTERNAL ANTARCTIC PLATEAU} 

\author{Franck Lascaux\supit{a}, Elena Masciadri\supit{a}, Susanna Hagelin\supit{a,b} and Jeff Stoesz\supit{a}
\skiplinehalf
\supit{a}INAF/Osservatorio Astrofisico di Arcetri, 5 L.go E. Fermi, 50125 Florence, Italy; \\
\supit{b}Uppsala Universitet, Department of Earth Sciences, Villav\"agen 16, S-752 36 Uppsala, Sweden
}

\authorinfo{Send correspondence to Franck Lascaux and Elena Masciadri.\\
E-mail: 
lascaux@arcetri.astro.it - masciadri@arcetri.astro.it.}

\begin{document} 
\maketitle 

\begin{abstract}
Mesoscale model such as Meso-Nh have proven to be highly reliable in reproducing 3D maps of optical turbulence 
(see Refs.~\citenum{Masciadri99a},~\citenum{Masciadri99b},~\citenum{Masciadri01},~\citenum{Masciadri04}) above 
mid-latitude astronomical sites.
These last years ground-based astronomy has been looking towards Antarctica. 
Especially its summits and the Internal Continental Plateau where the optical turbulence appears to be confined 
in a shallow layer close to the icy surface. 
Preliminary measurements have so far indicated pretty good value for the seeing above 30-35 m: 
0.36" (see Ref.~\citenum{Agabi06}) and 0.27" (see Refs.~\citenum{Lawrence04}, \citenum{Trinquet08}) at Dome C. 
Site testing campaigns are however extremely expensive, instruments provide only local measurements and 
atmospheric modelling might represent a step ahead towards the search and selection of astronomical sites thanks 
to the possibility to reconstruct 3D $C_N^2$ maps over a surface of several kilometers. 
The Antarctic Plateau represents therefore an important benchmark test to evaluate the possibility to 
discriminate sites on the same plateau.
Our group~\cite{Hagelin08} has proven that the analyses from the ECMWF global model do not 
describe with the required accuracy the antarctic boundary and surface layer in the plateau. 
A better description could be obtained with a mesoscale meteorological model. 
In this contribution we present the progress status report of numerical simulations (including the optical 
turbulence - $C_N^2$) obtained with Meso-Nh above the internal Antarctic Plateau. 
Among the topic attacked: the influence of different configurations of the model (low and high horizontal 
resolution), use of the grid-nesting interactive technique, forecasting of the optical turbulence during some 
winter nights.
\end{abstract}



\section{INTRODUCTION}
\label{sec:intro}  
The Internal Antarctic Plateau is, at present, a site of potential great interest for astronomical applications.
The extreme low temperatures, the dryness, the typical high altitude of the Internal Antarctic Plateau 
(more than 2500 m), joint to the fact that the optical turbulence seems to be concentrated in a thin surface 
layer whose thickness is of the order of a few tens of meters do of this site a place in which, potentially, 
we could achieve astronomical observations otherwise possible only by space.
\par
In spite of the exciting first results (see Refs.~\citenum{Lawrence04}, \citenum{Aristidi05},
\citenum{Trinquet08}) the uncertainties on the effective gain that astronomers might achieve from ground-based 
astronomical observation from this location still suffers from serious uncertainties and doubts that have been 
pointed out in previous work (see Refs.~\citenum{Masciadri06a}, \citenum{Geissler06}, \citenum{Hagelin08}). 
A better estimate of the properties of the optical turbulence above the Internal Antarctic Plateau can be achieved with both dedicated measurements done in simultaneous ways with different instruments and simulations provided 
by atmospheric models. 
Simulations offer the advantage to provide volumetric maps of the optical turbulence 
($C_N^2$) extended on the whole Internal Plateau and, ideally, to retrieve comparative estimates in a relative 
short time and homogeneous way on different places of the Plateau.
In a previous paper\cite{Hagelin08} our group performed a detailed analysis of the meteorological 
parameters from which the optical turbulence depends on, provided by the General Circulation Model (GCM) of the 
European Center for Medium-range Weather Forecasts (ECMWF).
In that work we quantified the accuracy of the ECMWF estimates of all the major meteorological parameters and, 
at the same time, we pointed out which are the limitations of the General Circulation Models. 
In contexts in which the GCMs fail, mesoscale models can supply more information. 
The latter are indeed conceived to reconstruct phenomena (such as the optical turbulence) that develop at a too 
small spatial and temporal scale to be described by a GCM.
In spite of the fact that mesoscale models can attain higher resolution than the GCM, thes parameters such as 
the optical turbulence are not explicitly resolved but are parameterized, i.e. the fluctuations of the 
microscopic physical quantities are expressed as a function of the corresponding macroscopic quantities 
averaged on a larger spatial scale (cell of the model). 
For classical meteorological parameters the use of a mesoscale model should be useless if GCMs such 
as the one of the ECMWF could provide estimate with equivalent level of accuracy. 
For this reason the Hagelin et al paper\cite{Hagelin08} has been a first step towards the exploitation of the 
mesoscale Meso-Nh model. 
We retrieved all what it was possible from the ECMWF analyses and we defined their limitations 
at the same time. We concluded that in the first 10-20 m, the ECMWF analyses show a discrepancy with respect to 
measurements of the order of 2-3 m.s$^{-1}$ for the wind speed and of 4-5 K for the temperature.
\par
Preliminary tests concerning the optimization of the model configuration and sensitivity to the horizontal and 
the vertical resolution with the Meso-Nh model have already been conducted by our team\citenum{Lascaux07} for 
the Internal Antarctic Plateau. 
In this paper we present further progress of that work. 
More precisely, we intend:
\begin{itemize}
\item To compare the performances of the mesoscale Meso-Nh model and the ECMWF General Circulation Model in 
reconstructing wind speed and absolute temperature (main meteorological parameters from which the optical 
turbulence depends on) with respect to the measurements. 
This analysis will quantify the performances of the Meso-Nh model with respect to the GCM from the ECMWF.
\item To perform simulations of the optical turbulence above Dome C employing different model configurations 
and compare the typical simulated thickness of the surface layer with the one measured by Trinquet et 
al~\cite{Trinquet08}~.
In this way we aim to establish which configuration is necessary to reconstruct correctly the $C_N^2$.
\end{itemize}
\begin{figure}
\begin{center}
\begin{tabular}{c}
\includegraphics[width=0.8\textwidth]{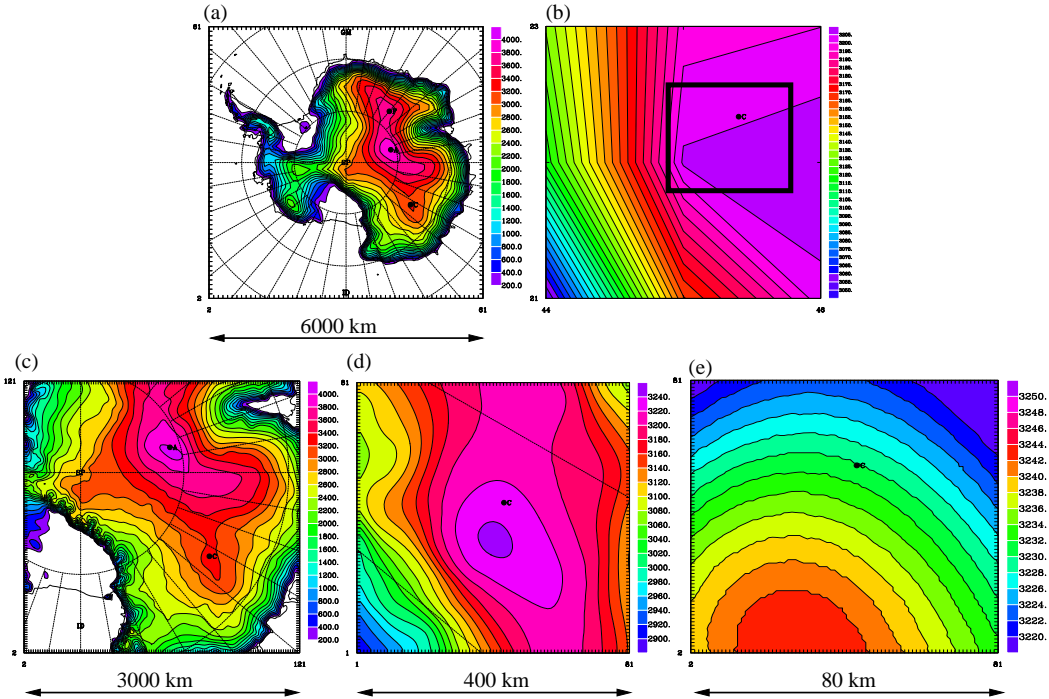}
\end{tabular}
\end{center}
\caption{\scriptsize Orography of Antarctica as seen by the Meson-Nh model (polar stereographic projection)
in (a) the monomodel simulation, with ${\Delta}X=100 km$; (b)
zoom of (a) above the Dome C area (the black square represents the same area as in (e));
(c) the dad model of the grid-nested simulation with ${\Delta}X=25 km$;
(d) the second domain of the grid-nested simulation with ${\Delta}X=5 km$;
(e) the innermost domain of the grid-nested simulation with  ${\Delta}X=1 km$.
The dot labelled 'C' is located at the Concordia Station. The dot labelled 'A' is the Dome A.}
{\label{fig:oro}}
\end{figure}
\section{MESO-NH MESOSCALE MODEL: NUMERICAL SET-UP} \label{sec:meso-nh}
Meso-Nh\cite{Lafore98} is the french non-hydrostatic mesoscale research model developed jointly 
by M\'et\'eo-France and Laboratoire d'A\'erologie.

It can simulate the temporal evolution of the three-dimensional atmospheric 
flow over any part of the globe. The prognostic variables forecasted by this model are the three cartesian components of the wind $u$, $v$, $w$, the dry potential temperature $\Theta$, the pressure $P$, the turbulent kinetic energy $TKE$.

The system of equation used is based upon the Lipps and Hemler~\cite{Lipps82} anelastic formulation allowing 
for an effective filtering of acoustic waves. 
A Gal-Chen and Sommerville\cite{Gal75} coordinate on the vertical and 
a C-grid in the formulation of Arakawa and Messinger\cite{Arakawa76} for the spatial digitalization is used.
The temporal scheme is an explicit three-time-level leap-frog scheme with a time filter\cite{Asselin72}.
The turbulent scheme is a one-dimensional 1.5 closure scheme\cite{Cuxart00} with the Bougeault and
Lacarr\`ere\cite{Bougeault89} mixing lenght.
The surface exchanges are computed in an externalized surface scheme (SURFEX) including different physical packages, among which ISBA\cite{Noilhan89} for vegetation.

Masciadri et al (see Refs.~\citenum{Masciadri99a}, \citenum{Masciadri99b})  implemented the optical turbulence 
package in order to be able to forecast 
not only the standard meteorological parameters with Meso-Nh, but also the optical turbulence ($C_N^2$ 3D maps) 
and all the astroclimatic parameters deduced from the $C_N^2$. We will refer to the Astro-Meso-Nh code to 
indicate this package. To compare simulations with measurements the integrated astroclimatic parameters are 
calculated integrating the  $C_N^2$ with respect to the zenith in the Astro-Meso-Nh-code. 
The parameterization of the optical turbulence and the reliability of the Astro-Meso-Nh model have been proved 
in successive studies in which simulations have been compared to measurements provided by different instruments 
(See Refs.~\citenum{Masciadri04},\citenum{Masciadri06b}). 
A dedicated calibration has been proposed and validated\cite{Masciadri01}~. 
\par
The atmospheric Meso-Nh model is conceived for research development and for this reason is in  
constant evolution. 
One of the major advantages of Meso-Nh that was not avalaible at the time of the Masciadri's studies is that it 
allows now the use of  the interactive grid-nesting technique\cite{Stein00}~.
This technique consists in using different imbricated domains with increasing horizontal resolutions with
mesh-sizes that can reach 10 meters. 
\par
We use in this study of the atmosphere above the Antarctica Plateau the same optical turbulence package,  implemented in the most recent version of the atmospheric Meso-Nh model. We list here down the differences in the model configuration that we implemented to do a step ahead in the research:

\begin{itemize}
\item We use a higher vertical resolution near the ground with respect to previous studies. 
We still work with a logarithimic stretching near the ground up to 3.5 km but we start with a first 
grid point of 2 m (instead of 50 m) with 12 points in the 
first hundred meters. This configuration has been allowed now thanks to a better description of the model 
pressure solver (available in the new atmospheric Meso-Nh version) and it is obviously preferable because it
permits to better quantify the turbulence contribution in the thin vertical slabs in the first hundred of meters.
Above 3.5 km the vertical resolution is constant and equal to ${\Delta}H=600$ m as well as in Masciadri's 
previous work. 
The maximum altitude is 22 kilometers.
\item The grid-nesting is implemented with 3 imbricated domains allowing a maximum horizontal resolution of 
1 km in a region around the telescope ($\sim$ 80 km $\times$ 80 km).
\item The simulation are forced at synoptic times by analyses from the European Center for Medium-range Weather 
Forecasts (ECMWF). This permits to perform a real forecast of the optical turbulence and it represents a 
step ahead with respect to the Masciadri et al.'s previous work. 
We highlight indeed, that in all the Masciadri's work the model has been used in a simple domain 
configuration permitting a quantification of the mean optical turbulence during a night and not really a 
forecast of the optical turbulence.
\end{itemize}

In spite of the fact that the orographic morphology is almost flat above Antarctica, we know that even a weak 
slope on the ground can be an important factor to induce a change in the wind speed at the surface in these 
regions. 
The physics of the optical turbulence strongly depend on a delicate balance between the wind speed and 
temperature gradients. In order to study the sensitivity of the model to the horizontal resolution we 
performed two sets of simulations with different model configurations. 

In the first configuration (that we will call {\it monomodel}) we use an horizontal resolution 
${\Delta}X=100$ km covering the whole Antarctic continent (Figure \ref{fig:oro}a,b). 
This permits to perform low time consuming simulations. Such a horizontal resolution has been also 
employed by Swain \& Gall\'ee\cite{Swain06} in a study on the boundary layer seeing done with regional atmospheric 
model MAR.

In the second configuration we used the grid-nesting technique to test the impact of the high-resolution 
on the optical forecasting. The grid-nested simulations involved three domains.
The biggest one has a 25 km mesh-size and covers all the Antactic Plateau with 120x120 points 
(Figure \ref{fig:oro}c).
The second one has a horizontal resolution of 5 km, 80x80 points and is centered above the Dome C 
(Figure \ref{fig:oro}d).
The innermost domain has a 1 km mesh-size, 80x80 points and is centered above the Concordia Station area near  
the Dome C (\ref{fig:oro}e).
\par
The use of high-resolution has one first major impact: the Dome C area is more fairly reproduced in the 
grid-nested simulation than in the low horizontal resolution simulation (Figure \ref{fig:oro}b,e).
The altitude above mean sea level of the Concordia Station with high resolution is around 3230 m, 
whereas it is around 3200 m with the low resolution grid.
More over in Fig. \ref{fig:oro}e we can observe that the Concordia Station is not located exactly in 
correspondence of the summit of Dome C but at $\sim$ 60 km far away with a slightly lower altitude 
($\Delta$h $\sim$ 15 m). 
Such a morphology can not be distinguished with the mono-model configuration.
\par 
In section 4 we will compare the typical thickness of the optical turbulence surface layer reconstructed by 
the model with the two different configurations with the measurements performed recently\cite{Trinquet08}  
to identify if the model is sensitive to this parameter (horizontal resolution) and in case 
this answer is positive, which configuration better matches with observations.
\par
In the next section we first start to compare results from a General Circulation model (ECMWF) and the two 
configurations from the mesoscale model Meso-Nh.
\par
All the simulations are initialized and forced every 6 hours with the analyses of the European Center of 
Medium-range Weather Forecast (ECMWF). 
\begin{figure}[ht]
\begin{center}
\begin{tabular}{c}
\includegraphics[width=0.9\textwidth]{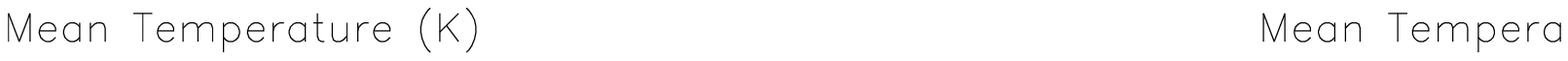}
\end{tabular}
\end{center}
\caption{\scriptsize Mean temperature profiles at the Concordia Station in the Dome C area over 47 winter nights (bold lindes) 
and the corresponding standard deviation (dashed lines). 
In black, from radiosoundings; in blue, from the ECMWF analyses; in green, from the Meso-Nh low horizontal 
resolution simulation; in red, from the Meso-Nh grid-nested simulation. 
On the left, profiles are plotted over 20 km. On the right, over the first kilometer.}
{\label{fig:mean_t}}
\end{figure}

\section{MESO-Nh SIMULATIONS IN WINTER ABOVE DOME C - METEOROLOGIC PARAMETERS} \label{sec:numerical}
The purpose of this section is to verify the performances of the mesoscale Meso-Nh model above the Internal 
Antarctic Plateau and to verify if such a mesoscale model can provide a better estimate of the atmospheric flow 
than a General Circulation model. 
\par
An important number of winter nights (47) were simulated with the Meso-Nh mesocale model. 
We analyze here the key meteorologic parameters from which the optical turbulence depends on: 
the temperature and the wind speed.
Both configurations (low horizontal resolution monomodel, and high horizontal resolution grid-nesting) are 
tested and evaluated.
All the simulations start at 00 UTC and are integrated for 12 hours. 
Simulations outputs at 12 UTC can be compared with measurements ({\it http://www.climantartide.it}) 
as well as to the analysis from the General Circulation model from the ECMWF.
Every 6 hours we force the simulations with the ECMWF analyses in order to avoid that the model diverge and/or 
correct the atmospheric flow as a function of the predictions at larger spatial scales.
\par
In this section a statistical study of the wind and temperature profiles at Concordia Station, Dome C, is 
performed.
The 47 nights have been selected in June, July and August 2005 and July 2006. 
For all the 47 nights selected, we respected the following criterion:
\begin{itemize}
\item A radiosounding is available  at the end of the simulation (at 12 UTC of the selected night) to perform 
comparisons between Meson-Nh outputs, ECMWF analysis and observations.
\item For the selected nights, the radiosoundings cover the longest path along the z-axis 
(perpendicolar to the ground) before to explode. 
It was impossible to collect in winter time 47 nights in which all the balloons reached the 20 km.
\end{itemize}
\subsection{Temperature profiles at Dome C}
\subsubsection{General behaviour of Meso-Nh}
The figure \ref{fig:mean_t} shows the mean temperature profiles at the Concordia Station computed over 47 winter nights from the two model configurations (low and high horizontal resolution), the ECMWF analyses and 
the radiosoundings.
All profiles have been interpolated on a regular 5 m vertical grid, in order to ease the comparison.
The mean temperature profiles are very similar over the entire free atmosphere (Figure \ref{fig:mean_t}, left).
Some light discrepancies appear on the first kilometer above the ground, where both the ECMWF analyses and the 
Meso-Nh model present a negative bias of 4-5 K up to 200 m.
At this altitude, the grid-nested simulation and the low resolution simulation give similar results and are not 
necessarily better than the ECMWF.
The temperature gradients provided by the two Meso-Nh configurations and the ECMWF analyses are not as 
pronounced 
as the one obtained with the radiosoundings. 
In the next section we will analyse the performances of ECMWF analyses and Meso-Nh model at the very surface 
level. i.e. the region in which the energetic fluxes budget between the ground and the atmosphere takes place. 
For this reason most of the ability for a model in reconstructing the correct atmospheric evolution near the 
ground relies on its ability in reconstructing the temperature of the surface.
\begin{table}[t]
\begin{center}
\begin{tabular}{|l|c||c|} \cline{2-3}
\multicolumn{1}{l|}{} &  Radiosondes & ECMWF \\
\hline
Windspeed (m.s$^{-1}$) & 4.02 ($\pm$ 2.55) & 6.51 ($\pm$ 2.51) \\
\hline
Temperature (K) & 212.90 ($\pm$ 7.64) & 216.64 ($\pm$ 5.83) \\
\hline
\end{tabular}
\end{center}
\caption{\scriptsize Mean values on 47 winter days at Concordia Station near Dome C of wind speed and
temperature at the surface level, for
radiosoundings, ECMWF analyses. Into brackets, the corresponding sigma.}
\label{tab:temp1}
\end{table}
\begin{table}[t]
\begin{center}
\begin{tabular}{|l|c||c|} \cline{2-3}
\multicolumn{1}{l|}{} &
\multicolumn{2}{c|}{Meso-Nh} \\ \cline{2-3}
\multicolumn{1}{l|}{} & 1-MOD & Grid-N \\
\hline
Windspeed (m.s$^{-1}$) & 4.23 ($\pm$ 1.77) & 3.98 ($\pm$ 1.95) \\
\hline
Temperature (K) & 214.92 ($\pm$ 4.64) & 214.50 ($\pm$ 4.97) \\
\hline
TKE (m$^2$.s$^{-2}$) & 0.39 ($\pm$ 0.30) & 0.35 ($\pm$ 0.33) \\
\hline
\end{tabular}
\end{center}
\caption{\scriptsize Mean values on 47 days at Concordia Station near Dome C of wind speed and
temperature at the surface level, for
Meso-Nh simulations: 1-MOD is for the low horizontal simulation with $\Delta$X=100 km and Grid-N is for
the high horizontal grid-nested simulation with $\Delta$X=1 km for the innermost model centered above the 
Dome C area.
Into brackets, the corresponding sigma.}
\label{tab:temp2}
\end{table}
\subsubsection{Surface temperature}
Tables \ref{tab:temp1}, \ref{tab:temp2}, \ref{tab:temp3} and \ref{tab:temp4} report the mean and median
temperature at the surface measured and simulated by the ECMWF analyses and the Meso-Nh model in the two
configurations (mono-model and grid-nesting).
\begin{table}[b]
\begin{center}
\begin{tabular}{|l|c|c|c||c|c|c|} \cline{2-7}
\multicolumn{1}{l|}{} &
\multicolumn{3}{c||}{Radiosondes} &
\multicolumn{3}{c|}{ECMWF} \\ \cline{2-7}
\multicolumn{1}{l|}{} &
Median & 75\% & 25\% & Median & 75\% & 25\% \\
\hline
Windspeed (m.s$^{-1}$) & 3.10 & 4.80 & 2.30 & 6.08 & 8.64 & 4.60 \\
\hline
Temperature (K) & 213.10 & 219.15 & 205.85 & 216.03 & 221.81 & 211.07 \\
\hline
\end{tabular}
\end{center}
\caption{\scriptsize Median values on 47 days and the corresponding quartiles at Concordia Station 
of wind speed and temperature at the surface level, for radiosoundings and ECMWF analyses and relatives 
percentiles.}
\label{tab:temp3}
\end{table}
\begin{table}[b]
\begin{center}
\begin{tabular}{|l|c|c|c||c|c|c|} \cline{2-7}
\multicolumn{1}{l|}{} &
\multicolumn{6}{c|}{Meso-Nh} \\ \cline{2-7}
\multicolumn{1}{l|}{} &
\multicolumn{3}{c||}{1-MOD} &
\multicolumn{3}{c|}{Grid-N} \\ \cline{2-7}
\multicolumn{1}{l|}{} &
Median & 75\% & 25\% & Median & 75\% & 25\% \\
\hline
Windspeed (m.s$^{-1}$) & 4.08 & 5.64 & 2.80 & 3.42 & 4.99 & 2.42 \\
\hline
Temperature (K) & 214.33 & 218.44 & 210.56 & 213.88 & 217.13 & 210.57 \\
\hline
TKE (m$^2$.s$^{-2}$) & 0.32 & 0.49 & 0.14 & 0.20 & 0.44 & 0.10 \\
\hline
\end{tabular}
\end{center}
\caption{\scriptsize Median values on 47 days and the corresponding quartiles at Concordia Station 
of wind speed and temperature at the surface level, for Meso-Nh simulations: 1-MOD is for the low horizontal simulation with $\Delta$X=100 km and Grid-N is for
the high horizontal grid-nested simulation with $\Delta$X=1 km for the innermost model centered at Concordia Station.}
\label{tab:temp4}
\end{table}
\\
One can see that the ECMWF analyses, as already reported in a previous paper of our team\cite{Hagelin08}~, are 
too warm at the surface (almost 4 K for the mean, and 3 K for the median) in winter with respect to the 
observations.
However, the surface temperatures (both mean and median) simulated by Meso-Nh after 12 hours are closer to the 
observations than the ECMWF analyses.
\par
The difference between ECMWF analyses and observed mean temperature is 3.74 K (2.97 K for the median).
The mean surface temperature in the low-resolution simulation is 2.02 K higher (1.23 K for the median) than in 
the observations.
It is only 1.60 K higher for the grid nested simulation (0.78 K for the median).
One can see that the surface temperature are even better retrieved with the use of the high horizontal resolution ($\Delta$X=1 km in the innermost domain) than the low horizontal resolution ($\Delta$X=100 km).
\subsection{Wind profiles at Concordia Station, Dome C area}
\begin{figure}[t]
\begin{center}
\begin{tabular}{c}
\includegraphics[width=0.9\textwidth]{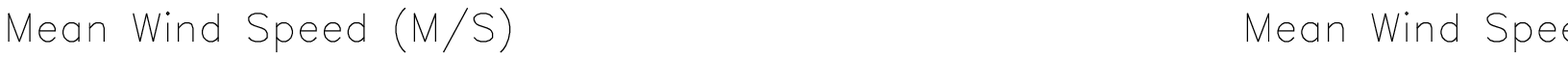}
\end{tabular}
\end{center}
\caption{\scriptsize Mean wind profiles at the Concordia Station in the Dome C area over 47 winter nights 
(bold lines)
and the corresponding standard deviation (dashed lines).
In black, from radiosoundings; in blue, from the ECMWF analyses; in green, from the Meso-Nh low horizontal
resolution simulation; in red, from the Meso-Nh grid-nested simulation. Units in m.s$^{-1}$.
On the left, profiles are plotted over 20 km. On the right, over the first kilometer.}
{\label{fig:mean_ws}}
\end{figure}
\subsubsection{General behaviour of Meso-Nh}
Figure \ref{fig:mean_ws} shows the mean wind during the 47 days, with the corresponding standard deviation. 
The highest mean altitude reached by the balloons is 10 km.
From the ground to 10 km, analyses and radiosoundings mean wind speed are well correlated.
Above 10 km the wind speed reconstructed by the ECMWF analyses is larger than that reproduced by Meso-Nh 
(monomodel and grid-nesting) (left of figure \ref{fig:mean_ws}). 
It is hard to say whether the ECMWF analyses or the Meso-Nh simulation is the best since no mean value from the 
observations is available at this altitude.
On the first kilometer above the ground, and especially around 150 m, it is well visible that Meso-Nh better 
reconstructs the strong wind shear than the ECMWF analyses.
The wind speed provided by the ECMWF analyses is a bit too weak (11 m.s$^{-1}$ instead of 13 m.s$^{-1}$ 
in the observations).
At the same altitude the Meso-Nh simulations give better results, with a mean wind profile perfectly 
correlated to the one measured by the radiosoundings.
The improvement is even better in the case of the high-resolution model.
The differences between low horizontal and high horizontal simulations is more important above 12 km, with 
an increase in intensity of the wind more important in the high-resolution simulation.
\subsubsection{Surface wind}
Tables \ref{tab:temp1}, \ref{tab:temp2}, \ref{tab:temp3} and \ref{tab:temp4} show the mean and median values 
of the wind speed at the first interpolated point of the profiles.
The mean wind speed in the ECMWF analyses is higher than the observed wind speed (6.51 m.s$^{-1}$ against 4.02 m.s$^{-1}$), thus a 
difference of 2.49 m.s$^{-1}$)(\footnote{Difference from the Hagelin et al paper\cite{Hagelin08} values are 
just due to the fact that in that paper, all the nights of the three months (June, July and August) are used 
while in this paper we simulated just 47 nights selected in two years.
The statistical sample is not therefore the same.}).
The difference in the median is of the same order of intensity: 3.02 m.s$^{-1}$.
Both low and high horizontal simulations reproduce better the surface wind than the ECMWF analyses.
With a mesh-size of 100 km in Meso-Nh, the difference in simulated mean wind speed ans observations is of 0.21 m.s$^{-1}$ 
(0.98 m.s$^{-1}$ for the median).
The grid nested simulations give even better results with a difference of 0.04 m.s$^{-1}$ only for the mean and 0.32 m.s$^{-1}$ 
for the median.
\begin{figure}[t]
\begin{center}
\begin{tabular}{c}
\includegraphics[width=0.8\textwidth]{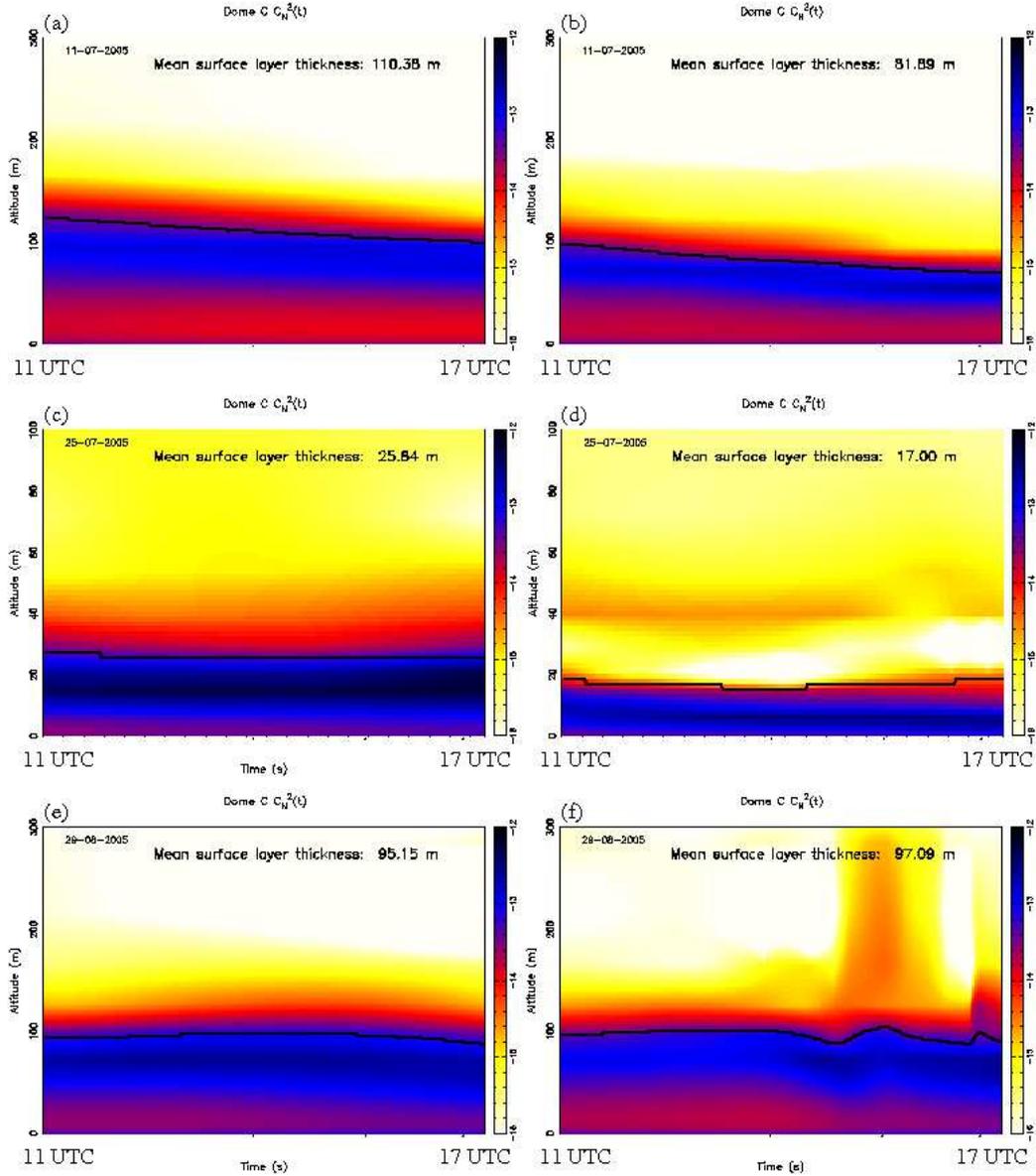}
\end{tabular}
\end{center}
\caption{\scriptsize Temporal evolution of the $C_N^2$ profile at Concordia Station obtained with Meso-Nh 
between 11 UTC and 17 UTC for (a) 2005 July 11, low horizontal resolution simulation, (b) 2005 July 11, 
grid-nested simulation,
(c) 2005 July 25, low horizontal resolution simulation, (d) 2005 July 25, grid-nested simulation,
(e) 2005 August 29, low horizontal resolution simulation and (f) 2005 August 29, grid-nested simulation.
The mean value of the computed mean height of the surface layer is written on the top-right of each graph
(see text for calculation details). The thick black line represent the evolution of the surface layer height.}
\label{fig:ot}
\end{figure}
\section{FORECAST OF OPTICAL TURBULENCE: FIRST NIGHTS}
In this section the optical turbulence package developed by our team\cite{Masciadri99a} and implemented 
in our Meso-Nh version of the model is used to perform the first simulations of some winter nights above 
the Antarctic Plateau. 
\par
All the 11 nights in the winter time (June, July and August) documented in Trinquet et al\cite{Trinquet08} were 
simulated.
We present here the first results obtained with Meso-Nh and concerning the height of the surface layer with 
the observations present in Trinquet et al\cite{Trinquet08}~.
\par
As an example, figure \ref{fig:ot} displays the temporal evolution of the $C_N^2$ profiles for three nights. 
The temporal windows starts at 11 UTC and ends at 17 UTC for each night.
All balloons have been launched between 14 and 15 UTC.
For each simulation, the average height of the surface layer is computed for the time period extended from 
11 UTC to 17 UTC.
The optical turbulence is indeed a parameterized parameter therefore it is a little hazardous to consider a 
precise time {\it forecasted time} as in the case of an explicitely resolved parameter such as the wind speed or 
the absolute temperature.
\par
In order to verify how much simulations match with measurements we compute the typical height of the surface 
layer using the same criterion as in Trinquet et al\cite{Trinquet08}~.
They defined the thickness $h_{sl}$ of the surface layer as the part containing 90\% of the first kilometer
boundary layer optical turbulence:
\begin{equation}
\label{eq:bl1}
\frac{ \int_{8m}^{h_{sl}} C_N^2(z)dz }{ \int_{8m}^{1km} C_N^2(z)dz } < 0.90
\end{equation}
This criterion is misleading, because it does not take into account the turbulence in the entire free 
atmosphere.
We propose to use a different criterion based on the computation of the total integrated $C_N^2$ over the 
whole 20 kilometers:
\begin{equation}
\label{eq:bl2}
\frac{ \int_{8m}^{h_{sl}} C_N^2(z)dz }{ \int_{8m}^{20km} C_N^2(z)dz } < 0.90
\end{equation}
The observed surface layer thickness for 11 winter nights from Trinquet et al\cite{Trinquet08} are reported on 
table \ref{tab:hl0}.
All mean values computed by Meso-Nh are reported in tables \ref{tab:hl1} (criterion described by Eq.\ref{eq:bl1})
and \ref{tab:hl2} (criterion described by Eq.\ref{eq:bl2}).
Different time intervals were chosen (between 12 UTC and 16 UTC, 11 UTC and 17 UTC, and 12 UTC and 18 UTC) 
for the mean calculation.
Independently from the time interval chosen for computing the mean, using the criterion of Trinquet et 
al\cite{Trinquet08} the grid-nested simulations give a mean surface thickness f around 48 m and the monomodel 
a thickness of 60 m.
Both configurations provide some higher mean thickness with respect to the observed one (36.9 m) but the 
grid-nested configuration is closer to observations ($\Delta$h $\sim$ 10 m).
In spite of the more expensive simulations, the grid-nested configuration seems to be necessary to better 
reconstruct the concentration of the turbulence in a thin layer near the surface.
\par
However, using our criterion (Eq.\ref{eq:bl2}), the mean thickness is now 64 m for the grid-nested simulation 
(75 m for the monomodel). 
This criterion provides, for both configurations (grid-nested and mono model) a surface layer $\sim$ 15 m 
thicker than what obtained with the criterion given in Eq.\ref{eq:bl1}.
This is far from being negligible for astronomical applications. 
The solution to overcome the turbulence surface layer with a tower might be put in serious discussion as a 
consequence of this result.
\begin{table}[t]
\begin{center}
\begin{tabular}{|c|c|} \cline{1-2}
Date &  Observed surface \\
& layer thickness (m)\\
\hline
13/06/05 & 23 \\
04/07/05 & 30 \\
07/07/05 & 21 \\
11/07/05 & 98 \\
18/07/05 & 26 \\
21/07/05 & 47 \\
25/07/05 & 22 \\
01/08/05 & 40 \\
08/08/05 & 30 \\
12/08/05 & 22 \\
29/08/05 & 47 \\
\hline
Mean     & 36.9 \\
\hline
\end{tabular}
\end{center}
\caption{\scriptsize Surface layer heights for 11 winter nights\cite{Trinquet08}~. Units in m.
The criterion used is the one from Eq.\ref{eq:bl1}.}
\label{tab:hl0}
\end{table}

\begin{table}[t]
\begin{center}
\begin{tabular}{|c|c|c|c|c|c|c|} \cline{1-7}
\multicolumn{1}{|c|}{Date} &
\multicolumn{3}{c|}{Surface thickness - Meso-Nh - Grid-N} &
\multicolumn{3}{c|}{Surface thickness - Meso-Nh - 1-MOD} \\
\hline
& {\scriptsize 11-17 UTC} & {\scriptsize 12-16 UTC} & {\scriptsize 12-18 UTC} & {\scriptsize 11-17 UTC} & {\scriptsize 12-16 UTC} & {\scriptsize 12-18 UTC}\\ \cline{2-7}
13/06/05 & 23.12 & 22.37 & 22.02 & 27.10 & 25.99 & 27.28 \\
04/07/05 & 26.18 & 25.43 & 25.77 & 42.14 & 42.55 & 41.01 \\
07/07/05 & 58.91 & 58.69 & 58.55 & 54.26 & 54.28 & 53.42 \\
11/07/05 & 81.89 & 81.36 & 77.62 &110.38 &110.01 &106.68 \\
18/07/05 & 54.15 & 53.54 & 51.80 & 84.87 & 84.88 & 81.73 \\
21/07/05 & 66.05 & 67.42 & 67.19 & 76.80 & 75.15 & 79.26 \\
25/07/05 & 17.00 & 16.56 & 17.16 & 25.84 & 25.62 & 25.86 \\
01/08/05 & 34.62 & 34.39 & 32.60 & 44.51 & 44.36 & 42.39 \\
08/08/05 & 40.66 & 41.75 & 38.52 & 64.09 & 69.42 & 60.36 \\
12/08/05 & 28.50 & 29.76 & 33.50 & 37.74 & 37.82 & 40.90 \\
29/08/05 & 97.09 & 98.26 & 95.04 & 95.15 & 96.51 & 93.47 \\
\hline
Mean     & 48.02 & 48.14 & 47.25 & 60.26 & 60.60 & 59.31 \\
\hline
\end{tabular}
\end{center}
\caption{\scriptsize Surface layer heights for 11 winter nights deduced  
from Meso-Nh computations using the criterion in Eq.\ref{eq:bl1}. The mean value 
is also reported.
Units in m.}
\label{tab:hl1}
\end{table}

\begin{table}[t]
\begin{center}
\begin{tabular}{|c|c|c|c|c|c|c|} \cline{1-7}
\multicolumn{1}{|c|}{Date} &
\multicolumn{3}{c|}{Grid-N} &
\multicolumn{3}{c|}{1-MOD} \\
\hline
& {\scriptsize 11-17 UTC} & {\scriptsize 12-16 UTC} & {\scriptsize 12-18 UTC} & {\scriptsize 11-17 UTC} & {\scriptsize 12-16 UTC} & {\scriptsize 12-18 UTC}\\ \cline{2-7}
13/06/05 & 23.70 & 23.07 & 22.57 & 27.82 & 26.54 & 27.72 \\
04/07/05 & 26.83 & 26.25 & 26.38 & 42.73 & 43.16 & 41.88 \\
07/07/05 & 59.93 & 59.47 & 59.54 & 55.88 & 56.13 & 55.32 \\
11/07/05 & 83.48 & 82.92 & 79.06 &112.07 &111.76 &108.33 \\
18/07/05 & 55.16 & 54.52 & 52.70 & 86.01 & 85.98 & 82.82 \\
21/07/05 &173.91 &194.67 &168.58 &156.87 &150.07 &170.65 \\
25/07/05 & 59.30 & 52.32 & 74.59 & 26.16 & 25.85 & 26.15 \\
01/08/05 & 35.08 & 34.87 & 33.03 & 45.45 & 45.33 & 43.31 \\
08/08/05 & 55.42 & 55.66 & 55.50 &142.80 &173.55 &135.57 \\
12/08/05 & 29.40 & 30.86 & 34.52 & 38.69 & 38.61 & 41.85 \\
29/08/05 & 98.39 & 99.53 & 96.20 & 95.86 & 97.31 & 94.26 \\
\hline
Mean     & 63.70 & 64.92 & 63.88 & 75.49 & 77.66 & 75.26 \\
\hline
\end{tabular}
\end{center}
\caption{\scriptsize Surface layer heights for 11 winter nights deduced
from Meso-Nh computations using the criterion in Eq.\ref{eq:bl2}. The mean value
is also reported.
Units in m.}
\label{tab:hl2}
\end{table}
The three nights presented here exhibit different turbulent conditions.
They show the ability of the model to react and predict the evolution of different surface layers.
During the first night (2005 July 11) the surface layer is constant in time, with a mean forecasted thickness of 
~82 m (with the grid-nested simulation, fig. \ref{fig:ot}b).
The observation for this night\cite{Trinquet08} gave a value of 98 m at 14:10 UTC.
The second night (2005 July 25) has a forecasted mean surface layer height lower 
(with the grid-nested simulation, fig. \ref{fig:ot}d): 17 m. 
The observed value was 22 m at 13:53 UTC. 
The third and last night displayed (2005 August 29) has a forecasted mean surface layer height of ~97 m 
(always with the grid-nested simulation, \ref{fig:ot}f).
The observed value for this night was 47 m at 14:47 UTC.
\par
In two of these three nights, the thickness of the surface layer retrieved by the model is well correlated with 
the observed one.
One can notice that for these two nights, the monomodel simulations give higher values than the grid-nested 
simulations.
The third night shows some interesting variability of the surface layer height, which are not present in the 
other nights and that is a signature of an evident temporal evolution of the turbulent energy distribution 
even in conditions of a pretty stratified atmosphere.
For this night Meso-Nh gives a thickness for the surface layer of around 97 m instead of the observed one (47 m).
\section{CONCLUSIONS}
In this paper we studied the performances of a mesoscale meteorological model, Meso-Nh, in reconstructing 
wind and temperature vertical profiles above Concordia Station, a site in the Internal Antarctic Plateau. 
Two different configurations were tested: monomodel low horizontal resolution, and grid-nesting high 
horizontal resolution.
The results were compared to the ECMWF General Circulation Model and radiosoundings.
\par
Several conclusions can be drawn:
\begin{itemize}
\item[(1)] We showed that near the surface, Meso-Nh retrieved better the wind vertical gradient than the ECMWF 
analyses, thanks to the use of a highest vertical resolution.
More over, the analysis of the first vertical grid point permits us to conclude that, as is, the Meso-Nh model 
surface temperature is closest to the observations than the ECMWF General Circulation Model which is too warm.
\item[(2)] The outputs from the grid-nested simulations are closer to the observations than the monomodel 
simulations.
This study highlighted the necessity of the use of high horizontal resolution to reconstruct a good 
meteorological field in Antarctica, even if the orography is almost flat over the Internal Antarctica 
Plateau.
The computations estimates from a previous study\cite{Swain06} are probably affected by the low horizontal 
resolution they used in their simulations.
\item[(3)] For what concerns the optical turbulence, both configurations predict a mean surface layer height 
higher than in the observations. 
However, it is always inferior at 100 m (when we used the criterion \ref{eq:bl1} of Trinquet et 
al\cite{Trinquet08}~). 
The configuration of Meso-Nh giving better estimate (closer to observations) is in grid-nesting mode.
\item[(4)] The results of Meso-Nh concerning the computation of the mean thickness of the surface layer are
not very dependent of the time interval used to average it.
This widely simplifies the analysis of simulations.
\item[(5)] The criterion used in Trinquet et al\cite{Trinquet08} appears to be misleading.
Indeed, it underestimates the typical thickness of the turbulent surface layer.
We propose the use of another critera (presented in the last section of the paper) instead. 
It gives a mean value higher of around 15 m for the eleven nights.
This result has an important implication for astronomical application.
It is indeed not so realistic to envisage a telescope placed above a tower of more than 50 m and the use of the 
Adaptive Optics remains the unique path to follow to envisage astronomical facilities at Dome C.
\end{itemize}
This optical turbulence study is based upon a little number of nights (11).
We need to extend it to a higher number of nights in order to have a more reliable statistical analysis of 
the results. 
\par
In this paper we did not present results concerning the seeing. 
We will focus our work ahead on this 
parameter, discriminating between two different partial contributions: the seeing in the free atmosphere, 
and the seeing of the surface layer.
We plan to make comparisons between Meso-Nh output and observations.

\acknowledgments     
The study has been carried out using radiosoundings from the Progetto di Ricerca "Osservatorio Meteo 
Climatologico" of the Programma Nazionale di Ricerche in Antartide (PNRA), {\it http://www.climantartide.it}.
ECMWF analyses are extracted from the MARS catalog, {\em http://www.ecmwf.int}.
This study has been funded by the Marie Curie Excellence Grant (FOROT) - MEXT-CT-2005-023878.

\end{document}